# IMPPAT 3.0: An updated FAIR database of phytochemicals and formulations of Indian Medicinal plants

Shanmuga Priya Baskaran[a,b,1], Ajaya Kumar Sahoo[a,b,1], Priyotosh Sil[a,b], Rahul Tiwari[a,b], Nikhil Chivukula[a,b], Sabrina Elsa Eapen[a], Geetha Ranganathan[a], Preeti Semwal[a], Areejit Samal[a,b,*]

[a] *The Institute of Mathematical Sciences (IMSc), Chennai 600113, India*

[b] *Homi Bhabha National Institute (HBNI), Mumbai 400094, India*

[1] S.P.B. and A.K.S. contributed equally to this work and should be considered as Joint-First authors.

[*]Corresponding author: asamal@imsc.res.in

**Address for correspondence:**

Areejit Samal

Computational Biology Group,

The Institute of Mathematical Sciences (IMSc),

CIT Campus, Taramani, Chennai 600113, India

Phone: +91-44-22543219

Email: asamal@imsc.res.in

## Abstract

Ayurvedic formulations represent a rich source of traditional knowledge, encompassing distinct therapeutic approaches, medicinal ingredients, and underlying principles of drug action. However, systematic integration of this knowledge with standardized modern biomedical terminology and resources remains limited. Here, we present IMPPAT 3.0, an updated version of the Indian Medicinal Plants, Phytochemistry And Therapeutics (IMPPAT: https://cb.imsc.res.in/imppat/) database that extends its scope by incorporating curated and harmonized information on Ayurvedic formulations from authoritative sources, while expanding its coverage of medicinal plants, phytochemicals, and therapeutic use terms. IMPPAT 3.0 further enriches plant and phytochemical information with annotations on clinical trials, patents, bioactivity, protein targets, spectral data, plant image links, and AyuCaRe reports. The compiled data is also organized as IMPPAT-KG, a knowledge graph that enables exploration of relationships among IMPPAT entities. We demonstrate the utility of the resource through a network medicine analysis of single herbal drugs (SHDs) reported for rheumatoid arthritis (RA), revealing significant proximity of their protein targets to RA-associated genes, and pinpointing of phytochemical pairs with potentially complementary effects on RA pathogenesis. In conclusion, by integrating traditional medicinal knowledge, IMPPAT provides a FAIR-compliant ethnopharmacological platform for systematic investigation of Indian medicinal plants and Ayurvedic formulations for drug discovery applications.

## 1. Introduction

For centuries, Indian medicinal plants have been used in traditional systems of medicine[1–5]. Ayurveda, in particular, has been practiced in India for thousands of years. Globally, around 80% of the population is estimated to use traditional medicine, while approximately 40% of pharmaceutical products have a natural product basis, highlighting the continuing relevance of traditional knowledge in modern healthcare[6].

Ayurvedic medicines are broadly based on single herbal drugs (SHDs) or combinations of multiple herbs, termed polyherbal formulations (PHFs). The latter are intended to achieve enhanced therapeutic effects compared to SHDs[7]. Ayurveda is also described as a preventive, predictive, and personalized system of medicine, making the scientific investigation of its principles particularly relevant to contemporary healthcare[8,9]. In this context, several studies have explored associations between Ayurvedic concepts, particularly Prakriti (phenotype-based human constitution), and fields such as genomics and biochemistry[8,10,11]. Establishing such scientific connections require systematic access to reliable, structured and Findable, Accessible, Interoperable, and Reusable (FAIR)[12]-compliant Ayurvedic knowledge.

The Ministry of AYUSH, Government of India, has developed authoritative resources such as The Ayurvedic Pharmacopoeia of India (API) and The Ayurvedic Formulary of India (AFI)[13], that provide expert-curated information on Ayurvedic formulations from traditional literature. Notably, we previously developed IMPPAT (https://cb.imsc.res.in/imppat/; Indian Medicinal Plants, Phytochemistry And Therapeutics) database, which is an extensive digital resource on Indian medicinal plants, their phytochemicals, and therapeutic uses[13,14]. Subsequently, several resources have been developed for similar or related purposes, including the Indian Medicinal

Plants Database by FRLHT/NMPB[16], GRAYU[17], OSADHI[18], BIMP[19], NEIST's herbal formulation database[20], and DRAVYA[21].

Here, we present IMPPAT 3.0, which expands the scope of previous versions (IMPPAT 2.0[14] and IMPPAT 1.0[15]) with increased coverage of Indian medicinal plants, their phytochemicals, and therapeutic uses. Importantly, IMPPAT 3.0 compiles information on Ayurvedic formulations including SHDs and PHFs from authoritative literature. Notably, IMPPAT 3.0 (similar to earlier version 2.0) provides FAIR[12]-compliant data through standardized entities linked to external resources, together with additional annotations on clinical trials, bioactivity, protein targets, patents, spectral information, plant image links, and AyuCaRe reports. The compiled data is also presented as a knowledge graph, IMPPAT-KG, enabling graph-based exploration. As a case study to highlight the utility of IMPPAT 3.0 database, we performed a network medicine-based analysis of SHDs indicated in Ayurveda for rheumatoid arthritis. In sum, IMPPAT 3.0 will serve as a bridge between traditional Ayurvedic knowledge and modern biomedical research, providing a basis for further scientific investigations in Ayurvedic biology.

## 2. Methods

### 2.1. Data curation and annotation exercise in IMPPAT 3.0

In this update, information on medicinal plants, phytochemicals, therapeutic uses, and formulations was curated and integrated from multiple authoritative and publicly available sources. Plant information was compiled from The API - part I, The AFI[22], National List of Essential AYUSH Medicines (NLEAM)[23] and sCentInDB[24], a database on essential oil profiles of Indian medicinal plants. Phytochemical and therapeutic use data were collected from RIMP volumes 1-31[1], sCentInDB[24], LOTUS[25] and COCONUT[26,27], while additional plant part-phytochemical associations were incorporated, wherever available. Traditional Ayurvedic formulation data, including the ingredient name, therapeutic uses, drug action properties, were curated from API and AFI published by the Ministry of AYUSH, Government of India. The curated plant, phytochemical, and therapeutic use data were standardized, and additional annotations were subsequently collected following the workflow developed in IMPPAT 2.0[14]. Additionally, medicinal plants and phytochemicals were mapped to clinical trial records in WHO ICTRP (https://www.who.int/tools/clinical-trials-registry-platform) and phytochemicals to patent records in PatCID[28], SureChEMBL[29] and BindingDB[30]. These records were identified based on the presence of the corresponding terms in the source databases. Phytochemicals were further linked to experimentally reported protein targets via information compiled in NPASS[31], ChEMBL[32] and BindingDB, bioactivities from NPASS and BindingDB, and spectral data from MassBank[33]. Plant entries were supplemented with Wikimedia Commons image links, and AyuCaRe reports. **Figure 1** provides an overview of the compiled data and annotations in IMPPAT 3.0.

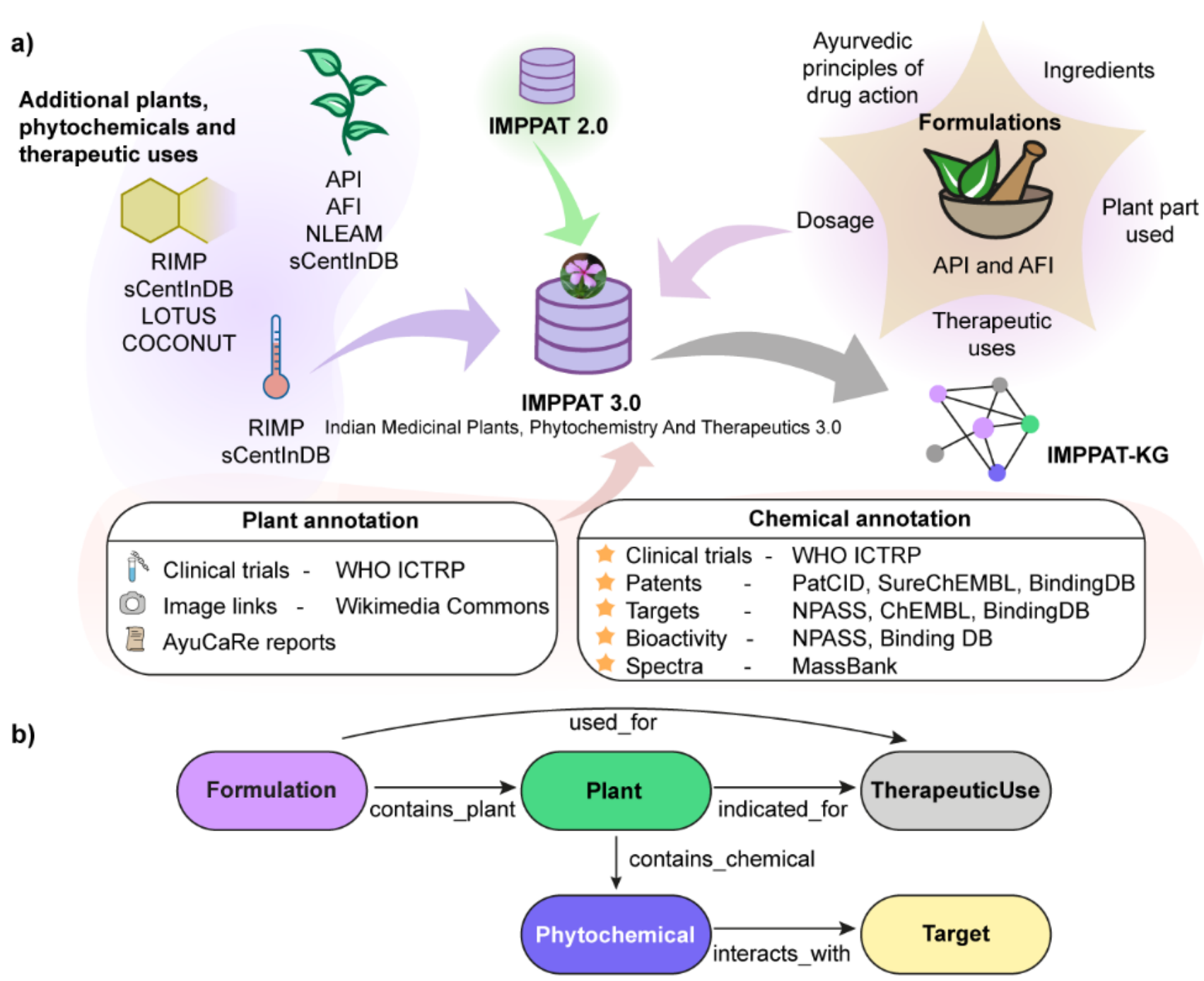

**Figure 1:** Summary of the compiled data and annotations included in IMPPAT 3.0. **(a)** Medicinal plant, phytochemical, therapeutic use, formulation data, and annotations along with their data sources. **(b)** Schematic representation of the IMPPAT knowledge graph (IMPPAT-KG) showing five node types and their relationships.

### 2.2. Ayurvedic formulations of Indian medicinal plants

Herbal formulations constitute an important class of Ayurvedic medicines and are widely used in Ayurvedic clinical practice[7]. In this study, the term formulation refers to both SHDs and PHFs. Data on SHDs were compiled from API part-I volumes 1-6, 8-9[4], while information on PHFs was obtained from AFI parts 1-4[3]. For each formulation, we curated the formulation name, ingredients (including plants and other ingredients), plant part, dosage, associated therapeutic uses, and drug action principles for SHDs namely, Rasa (taste), Guna (quality), Virya (potency), Vipaka (post-digestive effect), and Karma (action). The original records, some of which were documented in Sanskrit terminology, were translated into English using authoritative Ayurvedic books and reports published by the Ministry of Higher Education, Government of India, the Ministry of AYUSH, Government of India, and the WHO[3,4,34–37]. The translated plant ingredient names, drug action principles, and the therapeutic use terms were subsequently standardized and mapped to different resources using the workflow adopted in IMPPAT 2.0[14]. Each formulation was assigned a unique identifier. Further, a few formulations in API or AFI have the same name but differ in ingredients, therapeutic uses, and dosage, and such formulations were distinguished by adding numerical suffixes (e.g., Nīlī_1, Nīlī_2, Nīlī_3).

### 2.3. Construction of knowledge graph

Knowledge graphs (KGs) integrate and organize diverse types of information into a connected network, enabling the identification of new relationships and providing an efficient querying and reasoning framework[38]. Recent herbal databases such as HERB 2.0[39] and GRAYU[17], have employed KG framework to link plants, phytochemicals, therapeutic uses, formulations, and other associated entities, facilitating analyses such as the identification of molecular analogues and potential substitutes for threatened medicinal plants[17]. In this study, we constructed a KG comprising five entities, namely formulations, plants, phytochemicals, therapeutic use terms, and targets. The associations among these entities were derived from the curated information compiled within the IMPPAT 3.0 database. The KG included five types of relations: "indicated_for", linking plants to their therapeutic use terms; "contains_plant", linking formulations to the associated plants; "used_for", linking formulations to their therapeutic uses; "contains_chemical", linking plants to their associated phytochemicals; "interacts_with", linking phytochemicals to their target proteins (**Figure 1b**). Note that, only plant ingredients of formulations were represented in the KG, while non-plant ingredients were not included.

The KG was constructed based on Neo4j graph platform (https://neo4j.com/). The interactive interface for the KG was created using Knowledge-Graph-UI (KG-UI)[40], which leverages the Cypher language (https://neo4j.com/docs/cypher-manual/current/introduction/) to query the underlying KG and retrieve immediate neighbouring entities as well as the shortest path connecting any two entities (https://github.com/MaayanLab/Knowledge-Graph-UI/). The KG web interface in IMPPAT 3.0 is available as a separate tab (https://cb.imsc.res.in/imppat/kgraph/).

**2.4. Network proximity and separation analysis**

Rheumatoid arthritis (RA) is an autoimmune disease characterized by synovial inflammation, swelling, autoantibody production, bone destruction, and other systemic manifestations, with a higher prevalence among women. Several single herbal drugs (SHDs) are indicated for RA in Ayurveda. To investigate the potential scientific basis underlying the relationship between these SHDs and RA, we adopted a network medicine approach[41–43]. SHDs relevant to RA and their associated phytochemical information were obtained from IMPPAT 3.0. Phytochemicals were filtered based on drug-likeness and predicted bioavailability, and SHDs containing $\geq 10$ phytochemicals were included for further analysis (**Supporting Information**). Experimentally supported human protein targets of these phytochemicals were retrieved from IMPPAT. RA-associated genes were compiled by combining FDA-approved drug targets with gene associations supported by at least two disease-gene databases (**Supporting Information**; **Figure 3a**). Network analyses including proximity and separation measures were performed using NetMedPy[44] on an integrated human protein-protein interaction (PPI) network (**Supporting Information**). For each SHD, network proximity to the RA gene set was computed using the average shortest path length, and statistical significance was assessed via degree-preserving randomization strategy. Separation analysis was done by first selecting phytochemicals showing significant proximity (Z-score < 0; $p < 0.05$), and then for each SHD, pairwise network separation of the phytochemicals was computed to quantify topological complementarity (**Figure 3a**). Detailed methodology and implementation are provided in **Supporting Information**.

### 2.5. Web interface

The IMPPAT web platform has been updated to incorporate the newly curated information in version 3.0. Two major additions in the web interface are: (i) a Formulation section integrated into both the BROWSE and BASIC SEARCH interfaces, providing access to formulation-related information, and (ii) a KG module that enables exploration of the compiled data through an interconnected network. IMPPAT 3.0 is available at: https://cb.imsc.res.in/imppat. The database is developed with the help of MariaDB (https://mariadb.org), in-house HTML, PHP (https://www.php.net/), JavaScript, and jQuery (https://jquery.com/) scripts, open-source CSS framework Bootstrap 4.1.3 (https://getbootstrap.com/docs/5.3/), JSME Molecule Editor[45], JSmol (https://jmol.sourceforge.net/), Cytoscape.js[46], and jQuery plug-in DataTables. IMPPAT 3.0 is hosted on an Apache webserver running on Debian 9.4 Linux Operating system.

## 3. Results and discussion

### 3.1. Enhancement and expansion of IMPPAT

IMPPAT 3.0 represents the latest update of the database, with expanded coverage of Indian medicinal plants, phytochemicals, therapeutic uses, formulations, and their associated annotations (**Table 1**). This update adds phytochemical and therapeutic information for 144 new plants at the plant-part level, along with 398 phytochemicals and 469 therapeutic use terms. In IMPPAT 3.0, plants have additional annotations via links to 2,168 clinical trials, 187 AyuCaRe reports, and 3,328 Wikimedia Commons image links. Phytochemicals have additional annotations via links to 28,835 clinical trials, 14,349,740 patents, 1,943 human targets, 31,812 bioactivity records, and 12,817 spectral records (**Table 1**).

Notably, this update includes 1,576 Ayurvedic formulations, including 541 SHDs and 1,035 PHFs, together comprising 524 plants and 8,506 formulation-therapeutic use associations. IMPPAT 3.0 also compiles five Ayurvedic principles of action, namely, Rasa, Guna, Virya, Vipaka, and Karma for the SHDs. Among the plant ingredients in compiled formulations, *Butea monosperma* is most frequent in SHDs (6 formulations), while *Piper longum* and *Zingiber officinale* are most frequent in PHFs occurring in 397 and 373 formulations, respectively (**Figures 2a** and **S1a**). "Pain" is the most frequent therapeutic use for SHDs (201 formulations), whereas "cough" dominates for PHFs (129 formulations) (**Figures 2b** and **S1b**). Root and fruit are the predominant plant parts in SHDs and PHFs, respectively (**Figures 2c** and **S1c**). Leguminosae is the most represented plant family in SHDs, whereas Zingiberaceae dominates PHFs, followed by Piperaceae (**Figures 2d** and **S1d**). Although Leguminosae is the most represented family, followed by Compositae and Lamiaceae, the prominence of Zingiberaceae and Piperaceae in PHFs suggests the broad therapeutic utility[47–50] of associated herbs. Further, 335 plants are shared between SHDs and PHFs (**Figure 2e**), and the PHF Daśamūlāriṣṭa contains the highest number of plant ingredients (63). Purified sulphur is the most frequent non-plant ingredient (211 PHFs) (**Figure S1e**). Twelve plants are Endangered or Critically Endangered in SHDs and 16 plants in PHFs. Additional formulation statistics are provided in **Supporting Information** and in **Figure S2**.

In summary, IMPPAT 3.0 is a FAIR-compliant ethnopharmacology database of Indian medicinal plants and formulations with standardised, non-redundant terminology, providing one of the largest collections of Indian medicinal plants, phytochemicals and formulations, along with extensive annotations among comparable databases (**Table S1**).

**Table 1**: Comparison of database content and coverage between IMPPAT 2.0 and IMPPAT 3.0.

| **Database content** | **IMPPAT 2.0** | **IMPPAT 3.0** |
|---|---|---|
| Number of Indian medicinal plants | 4,010 | 4,154 |
| Number of phytochemicals | 17,967 | 18,314 |
| Number of standardized therapeutic use terms | 1,095 | 1,544 |
| Number of plant parts | 34 | 48 |
| Number of plant-part combinations | 17,944 | 18,571 |
| Number of plant-part-phytochemical combinations | 1,89,386 | 1,96,220 |
| Number of plant-part-therapeutic use combinations | 89,733 | 94,261 |
| Number of Ayurvedic formulations | NA | 1,576 |
| Number of plants in the formulations | NA | 524 |
| Number of clinical trials for plants | NA | 2,168 (for 509 unique plants) |
| Number of clinical trials for phytochemicals | NA | 28,835 (for 556 unique phytochemicals) |
| Number of patents associated with phytochemicals | NA | 14,349,740 (for 9297 unique phytochemicals) |
| Number of targets (with experimental evidence) | NA | 1,943 (for 2,944 unique phytochemicals) |
| Number of bioactivity records | NA | 31,812 (for 3,530 unique phytochemicals) |
| Number of spectral data records | NA | 12,817 records (for 1,157 unique phytochemicals) |
| Number of AyuCaRe reports | NA | 187 (for 226 unique plants) |
| Number of plants with Wikimedia Commons image links | NA | 3,328 |

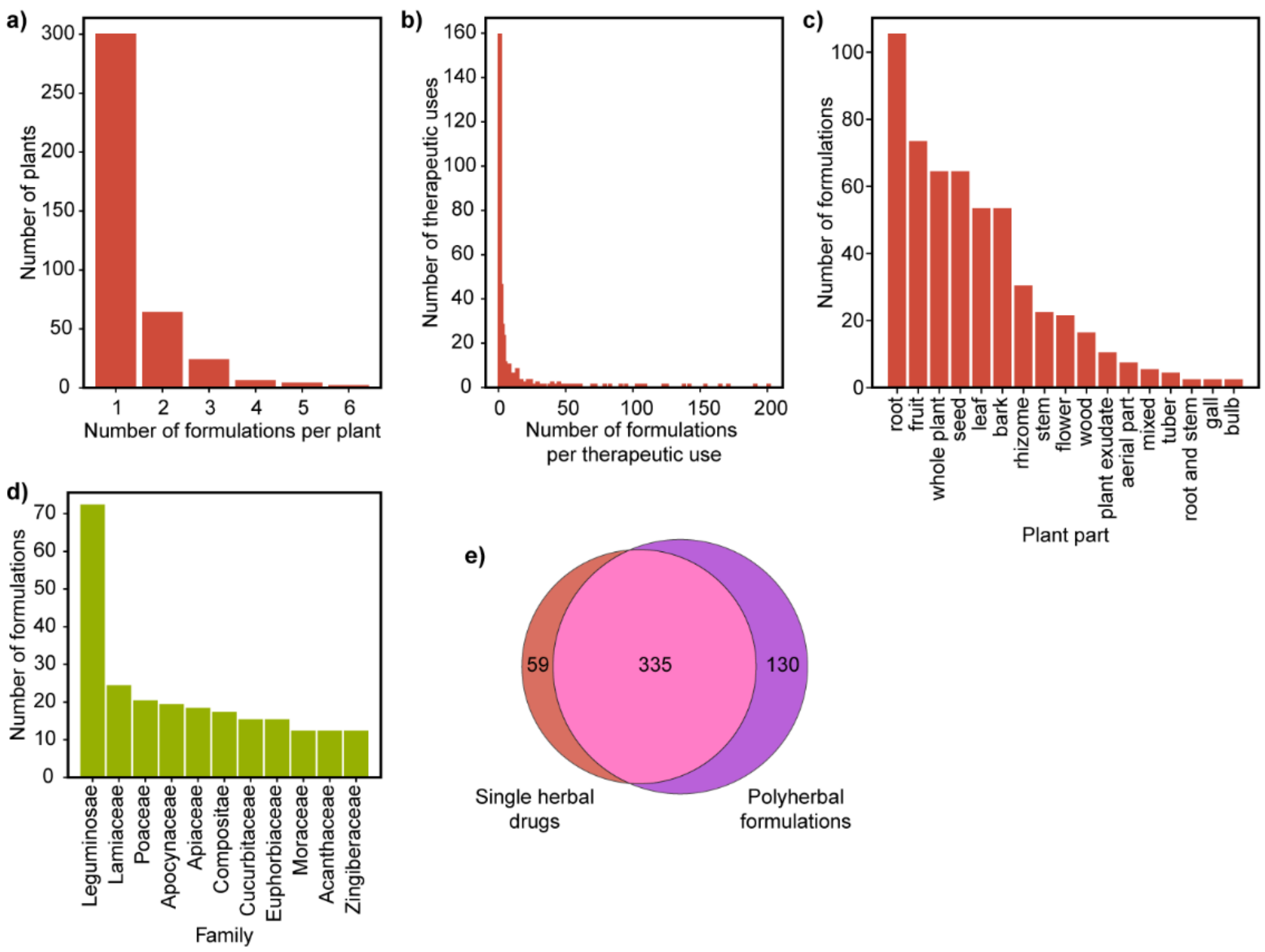

**Figure 2:** Statistics of curated SHDs in IMPPAT 3.0. **(a)** Distribution of plants by the number of associated formulations. **(b)** Distribution of therapeutic uses by the number of associated formulations. **(c)** Distribution of formulations by plant parts. **(d)** Number of formulations containing plants from the 11 most represented taxonomic families. **(e)** Overlap of plants between SHDs and PHFs.

### 3.2 IMPPAT-KG and web interface

IMPPAT-KG is constructed from compiled data in IMPPAT 3.0 (**Methods**). It comprises 27,513 unique nodes across five node types, including 4,154 plant nodes, 18,314 phytochemical nodes, 1,943 protein target nodes, 1,544 therapeutic use nodes and 1,558 formulation nodes. The KG also comprises 2,52,463 unique edges across five edge types. Among the edge types, "contains_chemical", linking plants and phytochemicals, was the most frequent, followed by "indicated_for", linking plants and therapeutic uses, while "used_for", linking formulations and therapeutic uses, is the least frequent. **Figure 1b** provides an overview of the nodes and their relationships in IMPPAT-KG**.**

IMPPAT-KG can be queried and visualized through a dedicated tab (https://cb.imsc.res.in/imppat/kgraph/), with searches based on node names or attributes such as common names and identifiers from the drop-down menu. This enables users to directly explore relationships among different entities within IMPPAT without downloading the underlying data. Beyond the KG, the IMPPAT web interface enables browsing of medicinal plants, phytochemicals, therapeutic uses, along with basic and advanced search options. Clinical trials, bioactivity records, protein targets, and patents have also been provided in the respective plant and phytochemical pages (**Figure S3**). IMPPAT 3.0 web interface additionally includes a dedicated formulation section allowing searches by formulation name, therapeutic use, and plant ingredients. The detailed formulation page provides information on ingredients, therapeutic uses, and documented Ayurvedic principles of drug action. A dedicated DOWNLOAD tab provides access to the IMPPAT 3.0 dataset for bulk download (**Figure S3**).

### 3.3. Case study I: Network proximity and separation analysis for RA-associated SHDs

A total of 40 SHDs (comprising 36 plant species) were identified for RA, of which 18 met the filtering criteria (**Methods**; **Supporting Information**; **Tables S2-S3**). These 18 SHDs comprised 597 unique phytochemicals, of which 206 had experimentally supported targets (835 target proteins) (**Tables S2-S4**). The RA disease-gene module consisted of 214 RA-associated genes, which was used for network computation (**Table S5**; **Figure 3a**).

Network proximity analysis showed that 17 out of 18 SHDs were significantly proximal to the disease module ($p < 0.05$), with Z-scores ranging from −1.140 to −4.288 (**Figure 3b**; **Table**

**S6**). Nirguṇḍī_1 exhibited the strongest association (Z = −4.288), indicating close network localization of its targets to the RA gene set. At the phytochemical level, 38 were significantly proximal to the disease-gene module (**Table S7**). Among SHDs with at least two proximal phytochemicals (14 SHDs), separation analysis identified 39 unique phytochemical pairs across 8 SHDs that potentially have complementary effects on RA (**Table S8**). Śuṇṭhī_1 (or Śuṇṭhī_2*)* showed the highest combinatorial potential (19 pairs). **Figure 3c** shows the top phytochemical pairs in terms of separation score ($s_{ij} > 0.25$).

Notably, benzaldehyde appeared in 11 unique pairs, frequently with eucalyptol, suggesting coordinated targeting of inflammatory, signaling, and oxidative stress pathways[51–53]. The taxifolin-hexanoic acid pair showed the highest separation score ($s_{ij}$ = 0.559). Other combinations, such as butyric acid-matairesinol ($s_{ij}$ = 0.517), indicate complementary modulation of gut microbiota-immune interactions and NF-κB-mediated inflammation[54,55]. Similarly, taxifolin-matairesinol pair reflects concurrent regulation of MAPK and NF-κB pathways[56,55]. These patterns indicate that phytochemicals within SHDs tend to occupy distinct yet disease-relevant network regions, supporting a systems-level mechanism where compounds modulate multiple pathological processes in RA in a complementary manner.

### 3.4. Case study II: Analysis for drug-substitute pairs

Abhava Pratinidhi Dravya describes the Ayurvedic practice of substituting an unavailable drug with an alternative one. In a curated list of 49 reported substitute pairs[57], 18 had both drugs present in IMPPAT 3.0 and were assessed for similarity across five Ayurvedic principles of action profiles (**Table S9**). Among the profiles, Rasa showed the highest concordance, matching in 17 of the 18 pairs, while six pairs matched across all five profiles. Details of the analysis are provided in the **Supporting Information**. This analysis suggests the potential relevance of Ayurvedic principles of action in Abhava Pratinidhi Dravya.

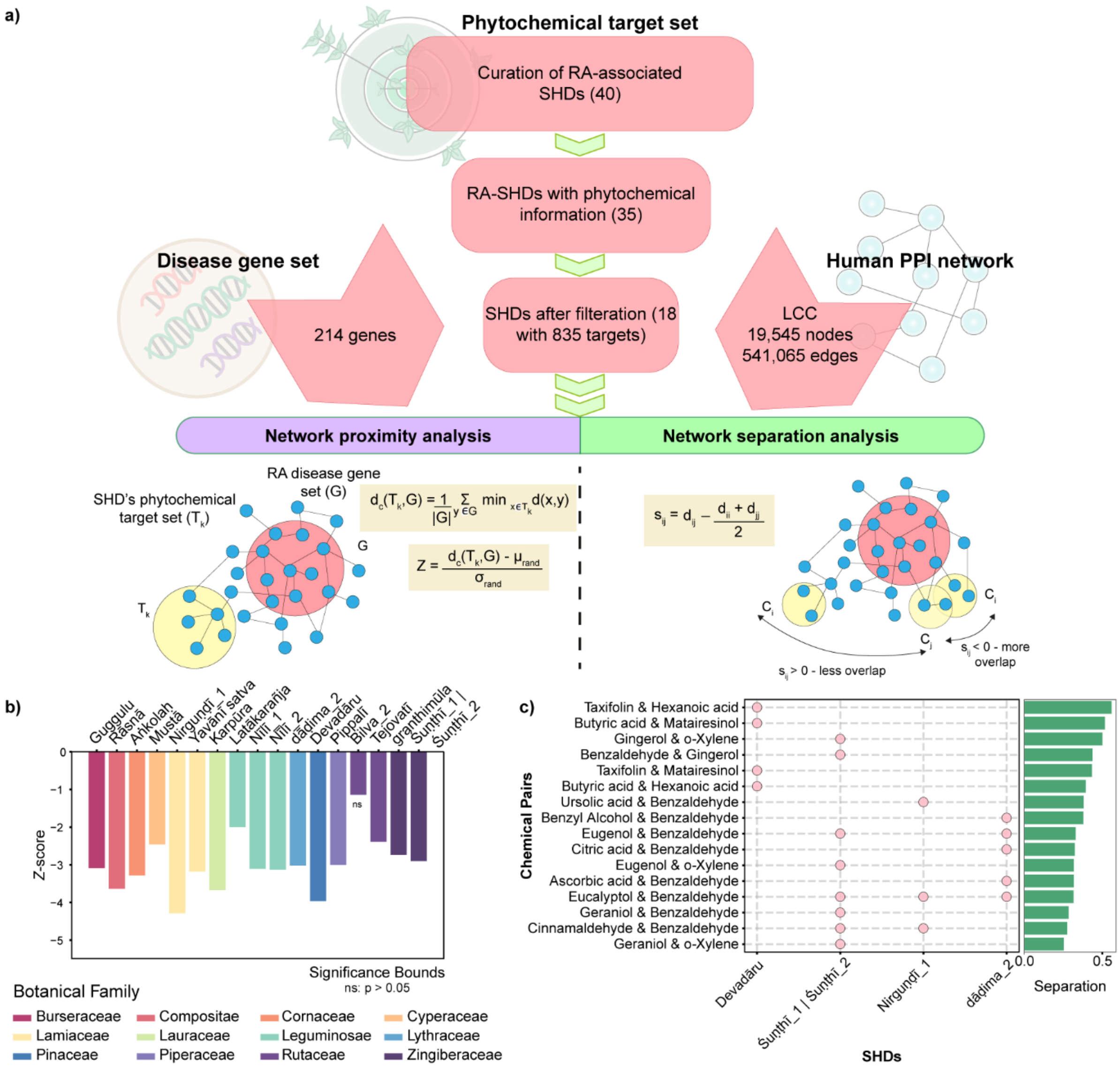

**Figure 3:** Network proximity and separation analysis of RA-associated SHDs in IMPPAT 3.0. **(a)** Workflow for network proximity and separation analysis of RA-associated SHDs. **(b)** Computed network proximity (Z-scores) of 18 SHDs to the RA-associated gene set, with negative values indicating closer than expected proximity. Bars are coloured by botanical family. **(c)** Top

phytochemical pairs with positive separation scores ($s_{ij} > 0$) among phytochemicals individually proximal to RA-associated genes. Dots indicate the presence of a given pair within an SHD, and horizontal bars denote separation scores.

**4. Conclusion**

Here, we present an updated IMPPAT database with additional medicinal plants, phytochemicals, therapeutic use terms, plant and phytochemical annotations, and 1,576 newly curated Ayurvedic formulations. We also developed IMPPAT-KG, a knowledge graph integrating the compiled data to facilitate exploration of relationships among IMPPAT entries. As a case study, we applied a network medicine approach to show that RA-associated SHDs exhibit statistically significant proximity to RA-associated genes and identified phytochemical pairs that could potentially exhibit complementary effects on RA pathogenesis. Furthermore, analysis of the Ayurvedic action profiles of SHDs among reported Abhava Pratinidhi Dravya revealed similarities across several of these profiles. In sum, this update extends the coverage and scope of IMPPAT by integrating FAIR-compliant ethnopharmacological information and annotations on medicinal plants and formulations.

Future development of IMPPAT could incorporate evidence from Prakriti-based studies[58], enabling computational approaches that better account for the principles and epistemology of Ayurveda[9]. Such integration could facilitate systematic analysis of Ayurvedic formulation patterns and their evolution in response to changing environmental and lifestyle conditions, similar to approaches explored in Traditional Chinese Medicine to characterize medication patterns and support novel formulation design[59–63]. Furthermore, despite advances in metabolomics, the chemical composition of medicinal plants and formulations remains incompletely characterized. Integrating Ayurvedic concepts with metabolomics could help prioritize and identify uncharacterized components. Together, these directions could strengthen IMPPAT as a platform for connecting Ayurvedic biology with modern biomedical research for drug discovery applications.

**Acknowledgements**

Areejit Samal acknowledges funding from the Department of Atomic Energy (DAE), Government of India through the project: Advanced Research and Education in Mathematical Sciences at IMSc [O.M. No. 2103/1/2025-R&D-II-DAE/IMSc/4885 dated 23.4.2026]. The funders have no role in the study design, data collection, data analysis, manuscript preparation, or decision to publish.

**Data availability**

The IMPPAT 3.0 database is accessible via the associated website: https://cb.imsc.res.in/imppat. The data associated with the database is available as flat files in the Download tab at https://cb.imsc.res.in/imppat/download. The compiled information in IMPPAT 3.0 is made available under a Creative Commons Attribution-NonCommercial-NoDerivatives 4.0 (CC BY-NC-ND 4.0) International License. Supporting Figures S1-S3 and Supporting Tables S1-S9 are available as the Supporting Information files.

**Author Contributions**

**Shanmuga Priya Baskaran** and **Ajaya Kumar Sahoo** contributed equally to this work and should be considered as joint-first authors. **Shanmuga Priya Baskaran:** Conceptualization, Data Curation, Formal Analysis, Methodology, Software, Visualization, Writing – original draft, Writing – review & editing; **Ajaya Kumar Sahoo:** Conceptualization, Data Curation, Formal Analysis, Methodology, Software, Visualization, Writing – original draft, Writing – review & editing; **Priyotosh Sil:** Conceptualization, Data Curation, Formal Analysis, Methodology,

Visualization, Writing – original draft; **Rahul Tiwari:** Conceptualization, Data Curation, Formal Analysis, Methodology, Visualization, Writing – original draft; **Nikhil Chivukula:** Software, Writing – original draft; **Sabrina Elsa Eapen:** Conceptualization, Data Curation; **Geetha Ranganathan:** Data Curation; **Preeti Semwal:** Data Curation; **Areejit Samal:** Conceptualization, Formal Analysis, Methodology, Supervision, Writing – original draft, Writing – review & editing.

## Conflicts of interest

The authors declare no competing financial interests.

## Supporting Information

Supporting Information consists of Supporting Figures S1-S3, Supporting Text S1-S3, and Supporting Tables S1-S9.

Supporting Figures

Statistics of curated Polyherbal formulations (PHFs) in IMPPAT 3.0 (Figure S1); Statistics of plant ingredients and Ayurvedic action principles of single herbal drugs (SHDs) and PHFs in IMPPAT 3.0 (Figure S2); Screenshots of IMPPAT website pages featuring the newly added data and functionalities (Figure S3).

Supporting Text

Additional statistics of Ayurvedic formulations in IMPPAT 3.0 (Supporting Text S1); Case study of SHDs used in Ayurveda to treat RA (Supporting Text S2); Case study for Abhava Pratinidhi Dravya (drug substitution) (Supporting Text S3).

Supporting Tables

Comparison of IMPPAT 3.0 with existing Indian medicinal plant and formulation databases (Table S1); Details of 40 Single herbal drugs (SHDs) associated with rheumatoid arthritis (RA) curated from IMPPAT (Table S2); Details of 18 SHDs used for network proximity and separation analysis (Table S3); Information on plant-phytochemical-target associations for 18 rheumatoid arthritis-associated SHDs (Table S4); Curated gene set associated with rheumatoid arthritis (Table S5); Computed network proximity of 18 SHDs to the rheumatoid arthritis disease-gene module (Table S6); Computed network proximity of 38 phytochemicals to the rheumatoid arthritis disease-gene module (Table S7); Computed network separation of phytochemical pairs that exhibited positive separation scores (Table S8); Details of 18 drug-substitute pairs considered for Abhava Pratinidhi Dravya case study (Table S9).

# Supporting Information

## IMPPAT 3.0: An updated FAIR database of phytochemicals and formulations of Indian Medicinal plants

Shanmuga Priya Baskaran[a,b,1], Ajaya Kumar Sahoo[a,b,1], Priyotosh Sil[a,b], Rahul Tiwari[a,b], Nikhil Chivukula[a,b], Sabrina Elsa Eapen[a], Geetha Ranganathan[a], Preeti Semwal[a], Areejit Samal[a,b,*]

[a] *The Institute of Mathematical Sciences (IMSc), Chennai 600113, India*

[b] *Homi Bhabha National Institute (HBNI), Mumbai 400094, India*

[1] S.P.B. and A.K.S. contributed equally to this work and should be considered as Joint-First authors.

[*]Corresponding author: asamal@imsc.res.in

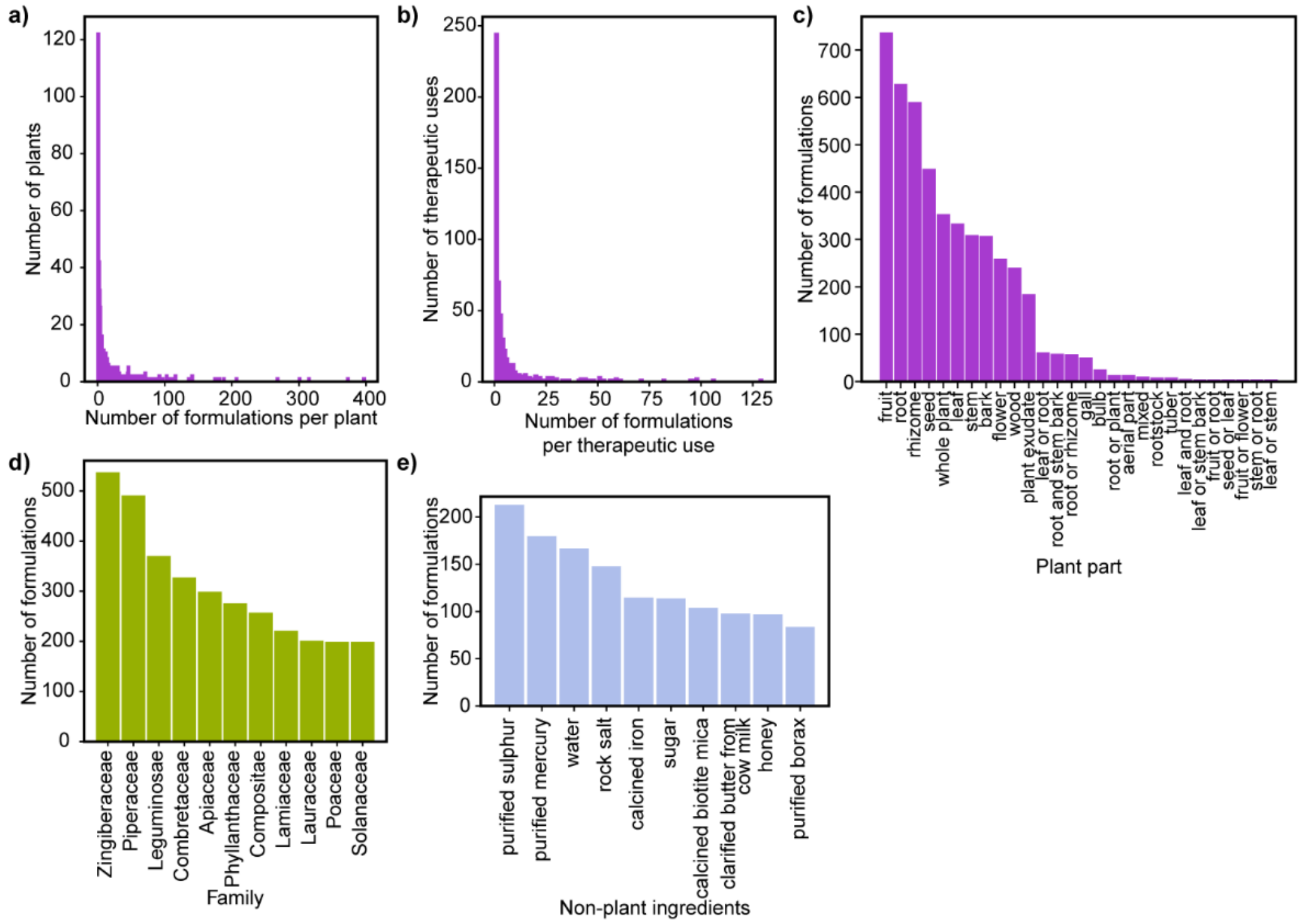

**Figure S1:** Statistics of curated Polyherbal formulations (PHFs) in IMPPAT 3.0. **(a)** Distribution of plants by the number of associated formulations. **(b)** Distribution of therapeutic uses by the number of associated formulations. **(c)** Distribution of formulations by plant parts. **(d)** Number of formulations containing plants from the 11 most represented taxonomic families. **(e)** Number of formulations containing the top 10 most frequently used non-plant ingredients.

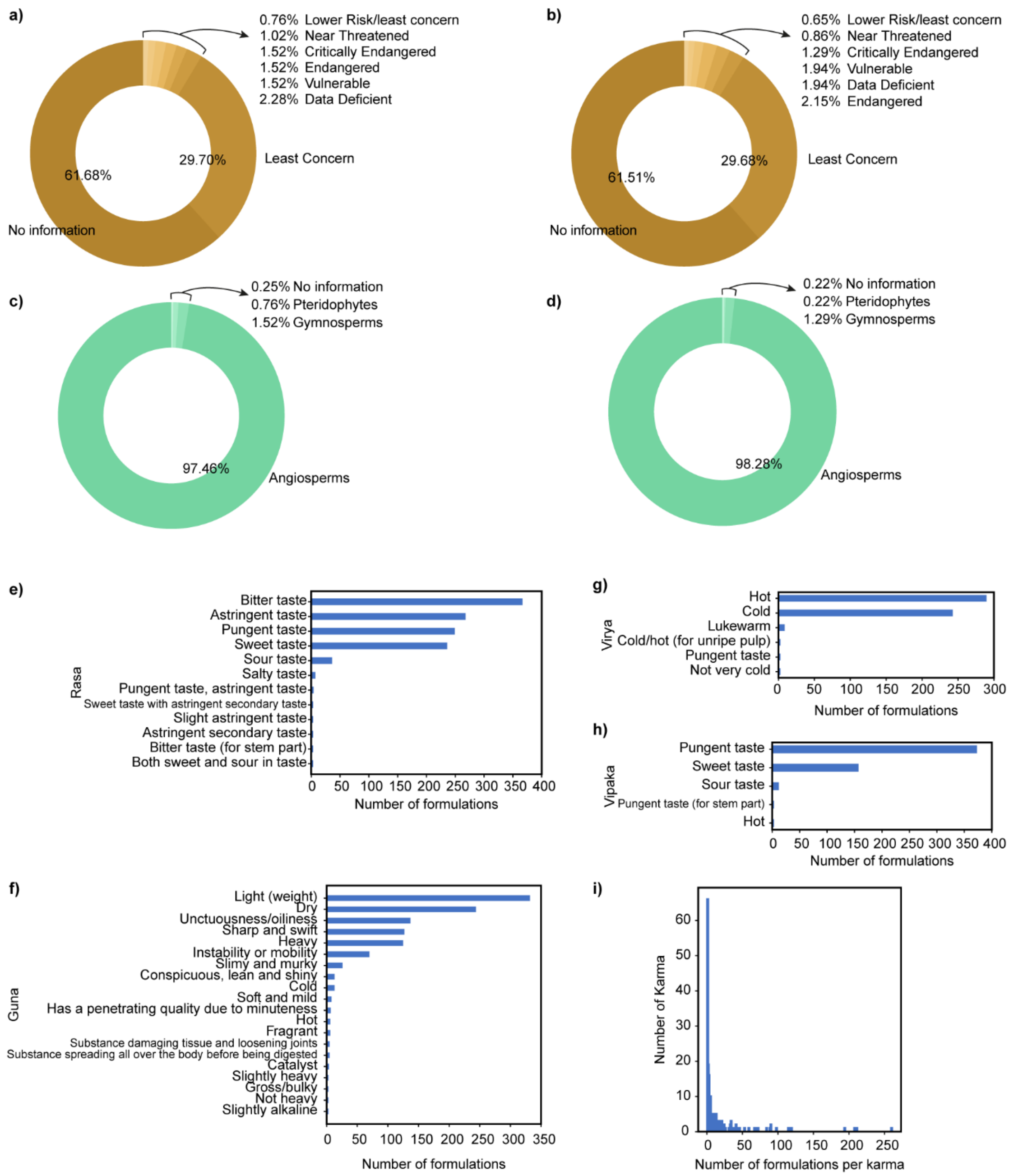

**Figure S2:** Statistics of plant ingredients and Ayurvedic action principles of single herbal drugs (SHDs) and PHFs in IMPPAT 3.0. Distribution of plants across IUCN Red List categories for **(a)** SHDs and **(b)** PHFs, respectively. Distribution of plants across groups for **(c)** SHDs and **(d)** PHFs, respectively. Distribution of the Ayurvedic action principles across SHDs for **(e)** Rasa, **(f)** Guna, **(g)** Virya, **(h)** Vipaka, and **(i)** Karma.

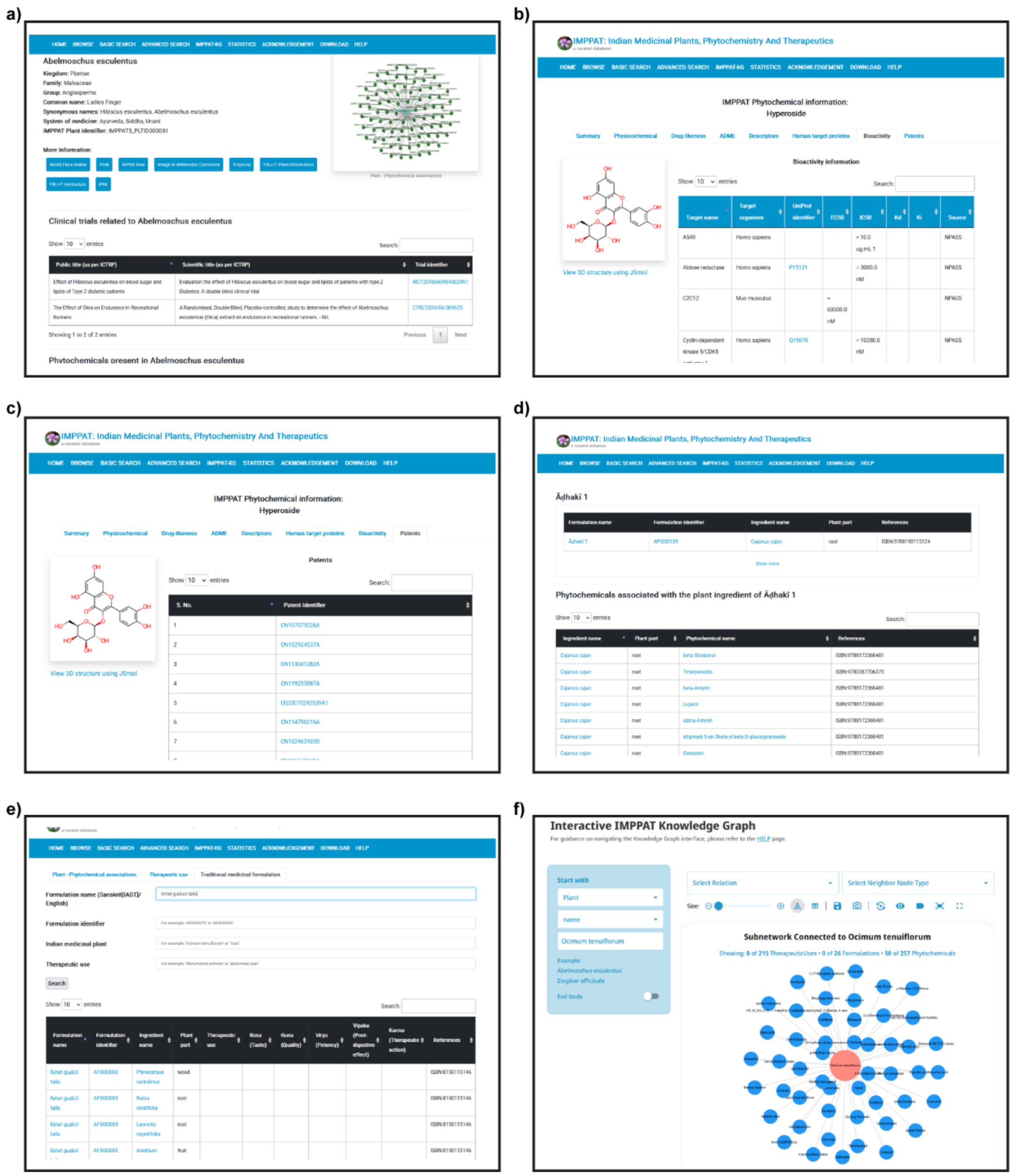

**Figure S3:** Screenshots of IMPPAT website pages featuring the newly added data and functionalities. **(a)** Clinical trial records for a plant, **(b)** bioactivity information, **(c)** patent records for a phytochemical, **(d)** detailed information for an SHD, **(e)** basic search interface for traditional medicinal formulations, with results displayed in tabular form, and **(f)** IMPPAT-KG visualization of a plant and a selected subset of its associated phytochemicals.

## S1. Additional statistics for Ayurvedic formulations in IMPPAT 3.0

There are 384 angiosperms, 6 gymnosperms, and 3 pteridophytes among the SHDs, compared with 457 angiosperms, 6 gymnosperms, and 1 pteridophyte among the herbs utilized in PHFs (**Figure S2c-d**). The distribution of Ayurvedic action principles among SHDs showed that bitter was the most frequently reported Rasa (**Figure S2e**). This predominance of bitter rasa was also observed among formulations associated with therapeutic conditions such as rheumatoid arthritis (RA) and diabetes mellitus, consistent with reports of the anti-inflammatory activity of bitter compounds[1]. In contrast, sweet rasa predominated among formulations associated with conditions such as fatigue and emaciation. Notably, drugs characterized by sweet rasa also predominantly exhibited sweet vipaka, suggesting their potential role in nourishment and strengthening[2,3]. It is important to note that Ayurvedic attributes such as Rasa, Guna, Virya and Vipaka may be represented by multiple terms rather than a single descriptor. Among the individual terms associated with these attributes, 'Light (weight)' was the most frequently reported Guna, 'pungent taste' was predominant in Vipaka, and 'Hot' is predominant in Virya (**Figure S2f-h**). Among Karma, 'Alleviates vitiated kapha' was the most frequently reported, occurring in 260 formulations (**Figure S2i**).

## S2. Case study of SHDs used in Ayurveda to treat Rheumatoid Arthritis (RA)

### S2.1. Systematic identification and characterization of SHDs for RA and their phytochemical composition

This case study focuses on SHDs used for the management of RA in Ayurveda as listed in The Ayurvedic Pharmacopoeia of India (API). Information on these SHDs was extracted from IMPPAT 3.0. The list of SHDs for rheumatoid arthritis (RA) was curated by screening SHDs documented with therapeutic use term 'rheumatoid arthritis' or 'āmavāta' or 'āmavātaḥ'. For each selected SHD, data on associated phytochemicals, their drug-likeness and predicted bioavailability, as well as corresponding protein targets, were retrieved from IMPPAT 3.0. Phytochemicals with no more than one Lipinski's rule violation and a predicted bioavailability score ≥ 0.50 were retained. To ensure statistical robustness and biological interpretability, only SHDs containing at least 10 phytochemicals were included in downstream analyses similar to our previous work[4].

**S2.2. Curation of phytochemical- and disease-associated human protein targets**

Human protein target information for phytochemicals associated with each SHD was retrieved from IMPPAT 3.0 for downstream analyses. The target associations had been curated based on experimental evidence from resources like NPASS[5], BindingDB[6], or ChEMBL[7]. For the analyses, target proteins were represented by their corresponding gene identifiers.

Disease-associated target sets for RA were compiled using an integrative, multi-source strategy. First, drugs relevant to RA were identified from DrugBank[8] by screening the Indication field using RA-specific keywords including "Rheum" and "Rheumatoid Arthritis," while excluding records associated with "Rheumatic Fever". Retrieved drugs were manually curated to retain only United States Food and Drug Administration (US FDA)-approved, non-withdrawn small molecules with clear therapeutic relevance. For each of the approved drug, therapeutic drug targets were curated from Therapeutic Target Database[9] (TTD), DrugCentral[10], and from highly cited articles[11–16]. Furthermore, multiple curated databases, including DisGeNet[17], GeneCards[18], Comparative Toxicogenomics Database[19] (CTD), and DISEASES[20] were utilized to obtain RA-associated genes. In the final RA-associated gene set, a gene was included if it satisfied at least one of the following criteria: (i) the gene was a known therapeutic target of an FDA-approved drug for RA; or (ii) the gene was reported in at least two of the four disease-gene association databases.

Gene identifiers were converted to their Entrez gene identifiers using the Syngo web server[21]. **Figure 3a** in the main text illustrates the workflow for data curation and subsequent analysis.

### S2.3. Network proximity and separation analysis

To investigate the systems-level therapeutic relevance of the selected SHDs for RA, a network-based analysis was performed using the Python package NetMedPy[22] on the human protein-protein interaction (PPI) network.

#### S2.3.1. Construction of protein-protein interaction network

A comprehensive PPI network was constructed for the computation of network proximity and separation by integrating multiple high-quality interactomes to maximize coverage. First, we incorporated the PPI network reported by Gan *et al.*, which comprised 18,505 nodes and 327,924 edges[23]. Second, we integrated the HI-union interactome from the Human Interactome Atlas[24] (https://interactome-atlas.org/). The interaction data in this resource were provided in Ensembl gene identifiers, which were converted to Entrez Gene identifiers using the Syngo web server[21]. After removing duplicate interactions and self-loops, the HI-union dataset comprised 9,053 nodes and 63,669 unique interactions. Finally, we incorporated the human PPI network curated by Ruiz *et al.*[25] (17,660 nodes and 387,626 edges), which integrates interaction data from multiple sources. All three networks were merged by taking the union of nodes and interactions, followed by removal of duplicate edges. The final integrated interactome consisted of 19,572 unique proteins and 541,092 unique interactions and was used as the background network for all subsequent analyses. Note that proteins in the PPI are represented by their corresponding gene identifiers.

#### S2.3.2. Network proximity analysis

To quantify the relationship between SHD-relevant protein targets and the RA-associated gene set, network proximity analysis was performed using NetMedPy (version 0.1.171)[22] in Python 3.11. All computations were conducted on the largest connected component (LCC; 19,545 nodes and 541,065 edges) of the integrated PPI network to ensure meaningful shortest-path distance calculations.

An SHD $H_k$ comprises a set of phytochemicals, where each phytochemical is associated with a set of protein targets. For a given SHD, all unique protein targets of its constituent phytochemicals were aggregated to define the SHD-specific target set $T_k$. The RA-associated gene set $G$ was constructed as described above. Network proximity was computed using the average shortest path length ($d_c$), as described by Guney *et al.*[26], in its asymmetric form. The closest distance between the SHD target set $T_k$ and the disease gene set $G$ was defined as:

$$d_c(T_k, G) = \frac{1}{|G|} \sum_{\{y \in G\}} min_{x \epsilon T_k} d(x, y) \qquad (1)$$

where $d(x, y)$ denotes the shortest-path distance between nodes $x$ (SHD targets) and $y$ (RA-associated genes) in the PPI network. Lower values of $d_c$ indicate closer topological proximity between SHD targets and RA-associated genes.

*Statistical significance assessment*

To evaluate whether the observed network proximity between the SHD target set $T_k$ and the RA-associated gene set was significantly smaller than expected by chance, a degree-aware null model was employed using NetMedPy[22]. Randomization was performed using the logarithmic binning strategy, which preserves the size and approximates the degree distribution of the original target set.

For each SHD, 1000 randomized target sets $T_k^{(r)}$ were generated. For each randomized set, the network proximity to the disease gene set $G$ was computed using the same asymmetric average shortest path length ($d_c$) definition as described above. This resulted in an empirical null distribution of proximity values.

The statistical significance of the observed proximity $d_c(T_k, G)$ was quantified using a *Z*-score:

$$Z = \frac{d_c(T_k,G) - \mu_{rand}}{\sigma_{rand}} \qquad (2)$$

where $\mu_{rand}$ and $\sigma_{rand}$ denote the mean and the standard deviation of the proximity values obtained from the randomized samples, respectively.

A one-tailed p-value was computed to assess whether the observed proximity was significantly smaller than expected under the null model. The p-value was defined as the fraction of randomized proximity values less than or equal to the observed value. Negative Z-scores together with $p < 0.05$ were considered indicative of statistically significant proximity between SHD targets and the RA-associated gene set.

**S2.3.3. Network separation analysis of phytochemicals**

To investigate the topological relationships among phytochemicals within each SHD, network separation analysis was performed using NetMedPy[22] on the LCC of the integrated PPI network. Network medicine studies have shown that the therapeutic effectiveness of multicomponent interventions can be interpreted in terms of the relative positioning of target modules within the interactome. In particular, prior work demonstrated that effective drug combinations tend to follow a characteristic topological pattern in which individual drug targets are proximal to the disease-associated gene module, yet remain topologically separated from each other, thereby enabling complementary modulation of disease-relevant pathways[27].

Motivated by this framework, we extended the same principle to phytochemicals within SHDs. Specifically, we hypothesized that phytochemical target proteins that are individually proximal to the RA-associated gene set but occupy distinct regions of the interactome may contribute complementary biological effects. To evaluate this, we quantified the network separation between phytochemical target sets following the identification of disease-proximal compounds.

*Selection of proximal phytochemicals*

For each SHD, network proximity analysis was first performed at the level of individual phytochemicals, as described above. Let $C_i$ denote the set of protein targets associated with phytochemical $i$. The proximity between $C_i$ and the RA-associated gene set $G$ was computed using the asymmetric average shortest path length. Phytochemicals whose target proteins exhibited statistically significant proximity to the disease gene set (Z-score < 0; $p < 0.05$) were retained for subsequent pairwise separation analysis.

*Pairwise network separation*

For each SHD, pairwise network separation was computed among the selected proximal phytochemicals. For two phytochemicals $i$ and $j$, with corresponding target sets $C_i$ and $C_j$, network separation was defined as:

$$s_{ij} = d_{ij} - \frac{d_{ii} + d_{jj}}{2} \qquad (3)$$

where $d_{ij}$ represents the mean shortest path distance between the target sets $C_i$ and $C_j$, and $d_{ii}$ and $d_{jj}$ denote the mean shortest path distances within each target set. Thus, the separation score compares the inter-module distance with the internal cohesiveness of the individual modules.

For phytochemicals with only one mapped target, the within-module distance is not defined due to the absence of internal node pairs. In such cases, the internal distance was set to zero to enable consistent computation of separation scores.

*Interpretation of separation score*

The separation score $s_{ij}$ captures the relative topological organization of the target set of phytochemicals within the interactome. Negative values of $s_{ij}$ indicate that the targets of two phytochemicals reside in overlapping network neighborhoods, suggesting shared or potentially redundant biological mechanisms. In contrast, positive values indicate that the target sets are located in distinct regions of the network, which may reflect complementary modes of action.

For each SHD, phytochemical pairs that: (i) individually exhibited significant proximity to the RA-associated gene set (Z-score < 0; $p < 0.05$), and (ii) showed separated network organization ($s_{ij} > 0$) were identified.

**S2.4. Characterization of SHDs, phytochemicals, and associated target and disease genes**

Querying the IMPPAT database resulted in the identification of 40 SHDs, corresponding to 36 plant species, documented for the management of RA in The Ayurvedic Pharmacopoeia of India (API) (**Table S2**). Among these, 35 SHDs (defined by unique plant-part combinations) had phytochemical information in the IMPPAT 3.0 database, each containing at least one associated phytochemical. To refine the dataset, phytochemicals were filtered based on drug-likeness and bioavailability criteria (see Section S2.1). The number of phytochemicals before and after filtering for each SHD is provided in **Table S2**. In addition, a minimum threshold of 10 phytochemicals per SHD was imposed for filtering to ensure sufficient phytochemical representation, improve statistical robustness, and biological interpretability. Following this filtration step, the dataset was reduced to 18 SHDs, representing a high-confidence subset for subsequent analysis (**Table S3**), corresponding to 16 plant species. Among these 18 SHDs, Śuṇṭhī_1 and Śuṇṭhī_2 have the same plant-part combination (*Zingiber officinale*, rhizome) and therefore, they are presented together in all tables and figures.

Across these 18 SHDs, a total of 597 unique phytochemicals were identified. The SHD-wise distribution of phytochemicals is provided in **Table S2**, while details about the associated phytochemicals are listed in **Table S4**. Notably, Śuṇṭhī_1 (or Śuṇṭhī_2) contributed the largest share, with 280 phytochemicals, whereas Nīlī_2 had the lowest representation, with 11 phytochemicals. Among the 597 phytochemicals, experimentally supported human protein targets were available for 206 compounds, corresponding to a total of 835 unique targets. Among the SHDs, Nīlī_2 exhibited the highest target coverage (519 targets), while Devadāru showed the lowest (95 targets). The SHD-wise distribution of targets is provided in **Table S3**, and associated details of the target proteins are listed in **Table S4**.

Disease-associated genes for RA were compiled based on the therapeutic targets of FDA-approved drugs and evidence from multiple databases (see Section S2.2). A total of 34 FDA-approved drugs were identified, corresponding to 18 curated therapeutic targets. Integration of these targets with disease-associated genes from multiple databases resulted in a non-redundant set of 214 RA-associated genes, defining the disease target space (**Table S5**).

### S2.5. Network proximity of SHD-derived targets to RA-associated genes

Network proximity analysis was performed to evaluate the relationship between each SHD and the RA-associated gene module (see Section S2.3). 17 out of 18 SHDs exhibited statistically significant proximity, as indicated by negative Z-scores and $p < 0.05$, indicating that their targets are topologically closer to the disease module than expected by chance (**Figure 3b** in main text; **Table S6**). Only Bilva_2 did not show statistically significant proximity as per p-value (although the corresponding Z-score is negative). Among all SHDs, Nirguṇḍī_1 demonstrated the strongest association, with a Z-score of −4.288 and $p < 0.001$. This observed proximity across most SHDs indicates that their constituent phytochemicals collectively target regions of the interactome relevant to disease-associated genes, supporting their potential role in modulating disease-related biological processes at a systems level.

**S2.6. Network separation of phytochemicals within SHDs**

Network proximity was first computed at the level of individual phytochemicals for the 206 compounds with available target information. Among these, 38 phytochemicals exhibited statistically significant proximity to the RA-associated gene set (**Table S7**). Thereafter, network separation was evaluated for phytochemical pairs within the 14 of 18 SHDs that contained at least two phytochemicals individually proximal to the disease module (see Section S2.3.3). In total, 39 unique phytochemical pairs across 8 SHDs (corresponding to 43 combinations across SHDs) satisfied the predefined criteria for separated network organization, suggesting complementary mode of action against RA (**Table S8**). Among these, Śuṇṭhī_1 or Śuṇṭhī_2 exhibited the highest number of pairs (19 pairs), whereas Bilva_2 and Latākarañja showed the lowest, with only one such pair, indicating variability in intra-herb combinatorial potential. **Figure 3c** in the main text shows the top phytochemical pairs in terms of separation score ($s_{ij} > 0.25$).

A notable observation was the frequent occurrence of benzaldehyde, which appeared in 15 of the 43 combinations, corresponding to 11 unique pairs, suggesting a central role in network-level complementarity. Benzaldehyde and its derivatives have been reported to possess anti-inflammatory and analgesic properties, including modulation of NF-κB signaling pathways[28,29]. This recurrent presence suggests that benzaldehyde may contribute substantially to the combinatorial therapeutic potential of SHDs.

Furthermore, four SHDs (dāḍima_2, Śuṇṭhī_1 or Śuṇṭhī_2, and Nirguṇḍī_1) shared a recurring phytochemical pair comprising benzaldehyde and eucalyptol, with a separation score of 0.315 (**Figure 3c**). While benzaldehyde is associated with anti-inflammatory and analgesic effects, eucalyptol has been implicated in the regulation of lipid peroxidation, a process linked to RA pathogenesis. The co-occurrence of these compounds suggests potential complementary targeting of distinct but related disease mechanisms[30].

The phytochemical pair taxifolin and hexanoic acid in Devadāru has the highest separation score of 0.559. A second phytochemical pair comprising butyric acid and matairesinol in Devadāru exhibited the second highest separation score (0.517) (**Figure 3c**). These compounds are associated with distinct yet complementary aspects of disease biology, with butyric acid implicated in the modulation of gut microbiota and immune homeostasis[31] and matairesinol linked to NF-κB mediated inflammatory pathways[32]. Their high network separation suggests potential complementary effects across multiple pathological processes. In addition, other phytochemical combinations also demonstrated separated network organization. For example, Taxifolin, which modulates inflammatory signaling pathways[33], and Matairesinol, acting via NF-κB-associated mechanisms[32], showed a separation score of 0.433, indicating coordinated targeting of inflammatory processes relevant to RA.

**S3. Case study for Abhava Pratinidhi Dravya (drug substitution)**

Abhava Pratinidhi Dravya is the Ayurvedic concept of using a substitute drug (pratinidhi dravya) when the original drug is unavailable (abhava dravya). To analyze the Ayurvedic principles of the documented substitute pairs, we obtained drug-substitute pairs from Venkatasubramanian *et al.*[34], which was originally compiled from several Ayurvedic literature sources, namely Bhavaprakasha Nighantu[35], Yogaratnakara, and Bhaishajya Ratnavali[36]. For each pair, we assessed the drug similarity across five Ayurvedic principle of action profiles, namely, Rasa, Guna, Virya, Vipaka, and Karma. A profile was considered matched if the two drugs shared at least one term. For example, the drug pair Kasha and Kusha have rasa profiles of 'Sweet taste, Bitter taste' and 'Sweet taste, Astringent taste', respectively, and were considered matched based on the shared term 'Sweet taste'.

Of the 49 reported drug-substitute pairs, 18 pairs had both drugs present among the SHDs curated in IMPPAT 3.0 and were evaluated for profile matching (**Table S9**). Note that drugs representing combinations of multiple drugs joined by 'and' were excluded from further analysis. Rasa matched in 17 of 18 pairs, showing the highest concordance among the five Ayurvedic principle of action profiles, consistent with its reported importance in influencing drug action[34]. Furthermore, six of the 18 substitute pairs exhibited a match across all five profiles, whereas Yashti and Dhataki matched only in virya. This analysis highlights the potential relevance of Ayurvedic principles of action in Abhava Pratinidhi Dravya. Although our analysis was based on a relatively small number of documented substitute pairs, a larger and more comprehensive dataset could provide further insights into the principles and framework underlying Ayurvedic drug substitution.